\documentclass[journal]{IEEEtran}
\usepackage{xcolor,soul,framed} 

\colorlet{shadecolor}{yellow}
\usepackage[pdftex]{graphicx}
\graphicspath{{../pdf/}{../jpeg/}}
\DeclareGraphicsExtensions{.pdf,.jpeg,.png}

\usepackage[cmex10]{amsmath}
\usepackage{algorithm}
\usepackage{algpseudocode}

\usepackage[hyperfootnotes=false]{hyperref}
\hypersetup{
 colorlinks=false,
 citecolor=Blue,
 linkcolor=false,
 urlcolor=false}

\usepackage{cite}
\usepackage{tablefootnote}
\usepackage{array}
\usepackage{mdwmath}
\usepackage{amssymb}
\usepackage{amsmath}
\usepackage{mdwtab}
\usepackage{eqparbox}
\usepackage{url}
\usepackage{xcolor}
\usepackage{comment}
\usepackage{footnote}
\usepackage{multirow}
\usepackage{graphicx}
\usepackage{caption}

\usepackage{float}
\usepackage{placeins}

\usepackage{url} 
\usepackage{array, makecell}

\usepackage{tikz}

\usetikzlibrary{fit,positioning}
\usetikzlibrary{bayesnet}
\usetikzlibrary{arrows}

\begin{document}
\bstctlcite{IEEEexample:BSTcontrol}
    \title{Predictive Models based on Deep Learning Algorithms for Tensile Deformation of AlCoCuCrFeNi High-entropy alloy}
    \author{Hoang-Giang Nguyen, Thanh-Dung Le,~\IEEEmembership{Member,~IEEE,}, Hong-Giang Nguyen and Te-Hua Fang

 \thanks{The authors acknowledge the support by the National Science and Technology Council, Taiwan, under grant numbers NSTC 110-2221-E-992-037-MY3. \textit{(Corresponding author: Te-Hua Fang)}. }

\thanks{Hoang-Giang Nguyen is with the Department of Mechanical Engineering, National Kaohsiung University of Science and Technology, Kaohsiung 807, Taiwan. He is also with the Faculty of Engineering and Technology, Kien Giang University, Kien Giang Province, Vietnam (Email: nhgiang@vnkgu.edu.vn ). }

\thanks{Thanh-Dung Le is with the Department of Electrical Engineering, \'{E}cole de Technologie Sup\'{e}rieure, University of Qu\'{e}bec,  Montr\'{e}al, Qu\'{e}bec, Canada (Email: dung.le@inrs.ca).}

\thanks{Hong-Giang Nguyen is with the Department of Academic and Students` Affairs, Hue University, Hue City, 49000, Vietnam (Email: giangnh@hueuni.edu.vn).}

\thanks{Te-Hua Fang is with the Department of Mechanical Engineering, National Kaohsiung University of Science and Technology, Kaohsiung 807, Taiwan (Email: fang@nkust.edu.tw 
).}
 
}

\markboth{IEEE, VOL., NO., 2024.
}{Hoang-Giang Nguyen \MakeLowercase{\textit{et al.}}: Predictive Models based on Deep Learning Algorithms for Strain Deformation from Nano-material}

\maketitle

\begin{abstract}

High-entropy alloys (HEAs) stand out between multi-component alloys due to their attractive microstructures and mechanical properties. In this investigation, molecular dynamics (MD) simulation and machine learning were used to ascertain the deformation mechanism of AlCoCuCrFeNi HEAs under the influence of temperature, strain rate, and grain sizes. First, the MD simulation shows that the yield stress decreases significantly as the strain and temperature increase. In other cases, changes in strain rate and grain size have less effect on mechanical properties than changes in strain and temperature. The alloys exhibited superplastic behavior under all test conditions. The deformity mechanism discloses that strain and temperature are the main sources of beginning strain, and the shear bands move along the uniaxial tensile axis inside the workpiece. Furthermore, the fast phase shift of inclusion under mild strain indicates the relative instability of the inclusion phase of HCP. Ultimately, the dislocation evolution mechanism shows that the dislocations are transported to free surfaces under increased strain when they nucleate around the grain boundary. Surprisingly, the ML prediction results also confirm the same characteristics as those confirmed from the MD simulation. Hence, the combination of MD and ML reinforces the confidence in the findings of mechanical characteristics of HEA. Consequently, this combination fills the gaps between MD and ML, which can significantly save time human power and cost to conduct real experiments for testing HEA deformation in practice.

\end{abstract}


\IEEEpeerreviewmaketitle

\section{Introduction}

Molecular dynamics (MD) is a simulation method that investigates the physical motions of atoms and molecules in N-body systems \cite{cazorla2017simulation, leimkuhler2015molecular, paesani2016getting}. It computes the paths of these particles by solving Newton's equations of motion \cite{hu1992direct, leal1979motion}. MD is used to get fundamental insights into the deformation mechanisms of metals and alloys \cite{bomarito2015atomistic, osetsky2010atomic, osetsky2003atomic, davoodi2016molecular, bahramyan2016molecular, murty2019high, zhang2014microstructures, cantor2004microstructural}. High entropy alloys (HEAs) represent a unique and attractive group of multi-component alloys comprising more than five primary elements in varying atomic percentages \cite{nguyen2023cyclic, bahramyan2020determination, singh2011decomposition, nadutov2015effect, hemphill2012fatigue}. One of the HEAs that has been extensively studied for its outstanding mechanical properties is Al\textsubscript{x}CrCoFeCuNi, with \( x \) (the molar ratio in the alloy) ranging from 0 to 3 \cite{pickering2015fine, doan2022effects, wu2006adhesive, tsai2009deformation, wang2012effects, roy2014fracture, xie2016molecular, xie2013alcocrcufeni, shaysultanov2013phase, tong2005microstructure, wang2009tensile, xu2015nanoscale}. 

The AlCoCrCuFeNi HEA is one of the most well-known and extensively researched HEAs; studies have focused on phase formation, microstructure, and mechanical properties \cite{kim2019selective, fang2018deformation, haghdadi2020hot, nguyen2023mechanics, tung2007elemental}. The number of related studies has expanded recently due to the development of a newly designed AlCoCrCuFeNi HEA based on structural performance optimization \cite{karlsson2019elemental, manzoni2016path}. 

In order to fully comprehend the deformation mechanism of the material, it is imperative to take into account the existence of the hexagonal close-packed (HCP) phase from the onset as an inclusion in the substrate. The HCP phase typically arises in the crystal structure during the deformation of the material. Defects typically manifest during processes like tension, compression, indentation, imprinting, and cutting of materials. Numerous studies have explored the impact of temperature or strain rate on the mechanical properties of materials \cite{nguyen2022temperature, lu2019research, chen2019evaluation}.

MD simulations have been essential in improving our understanding of deformation, phase transitions, and stress, particularly at the nanoscale, when used in conjunction with experimental studies. This is important since it is difficult for experiments to thoroughly explore material phenomena at this scale \cite{almotasem2017tool, wang2016investigations, nguyen2022effects}.

To the best of our knowledge, there has yet to be an atomic-scale exploration of the tensile behavior of AlCoCrFeCuNi alloys, considering different temperatures, strain rates, and grain sizes, aiming to predict outcomes using machine learning. Understanding the impact of temperature, strain rate, and grain size on mechanical properties and deformation mechanisms requires a comprehensive investigation of these alloys under varied conditions. This study utilizes MD simulations on AlCrCoFeCuNi alloys with FCC structures to examine their behavior under uniaxial tensile stress at different temperatures, strain rates, and grain sizes.

In this investigation, we explore how temperatures, strain rates, and grain sizes impact the deformation mechanism of polycrystalline AlCoCrCuFeNi HEA using MD simulation and machine learning support. 

Technically, firstly, we will conduct the MD simulation to obtain the atomic models of this AlCoCrCuFeNi HEA, including assigning grain coloring, designing the workpiece, and changing the polycrystalline structures. Hence, we will get the simulation results for the mechanical properties of AlCoCrCuFeNi HEA from the MD simulation under the effect of different scenarios of grain size, temperature, and strain rate changes. Those simulation results will be compared with existing studies (simulation and experiments), and then those results will be evaluated and concluded as complying with the theory before we move to the next step. 

Secondly, from the validated simulation results from the MD step, we will continue to adapt the ML algorithm to predict the strain deformation of the AlCoCrCuFeNi HEA under the different mechanical setups. Specifically, we experiment with  6 different ML algorithms. Then, evaluate the best ML predictive model based on the evaluation metrics, including Mean Absolute Percentage Error (MAPE), Mean Absolute Error (MAE), Root Mean Square Error (RMSE), and R-Square (R2). We also analyze the important of each material character as the contribution to the strain deformation. 

Consequently, by the combination of MD and ML, the study unveils the tensile properties of the plastic deformation in AlCoCrCuFeNi HEA, determining its dependence on various conditions. These findings contribute to a deeper understanding of the mechanical capabilities of AlCoCrCuFeNi HEA.

\section{Materials and Methods}
\subsection{MD Simulation and Model}
\label{sec:data_NKUST}
\begin{figure}[!htp]
	\centering
	\includegraphics[scale=0.06]{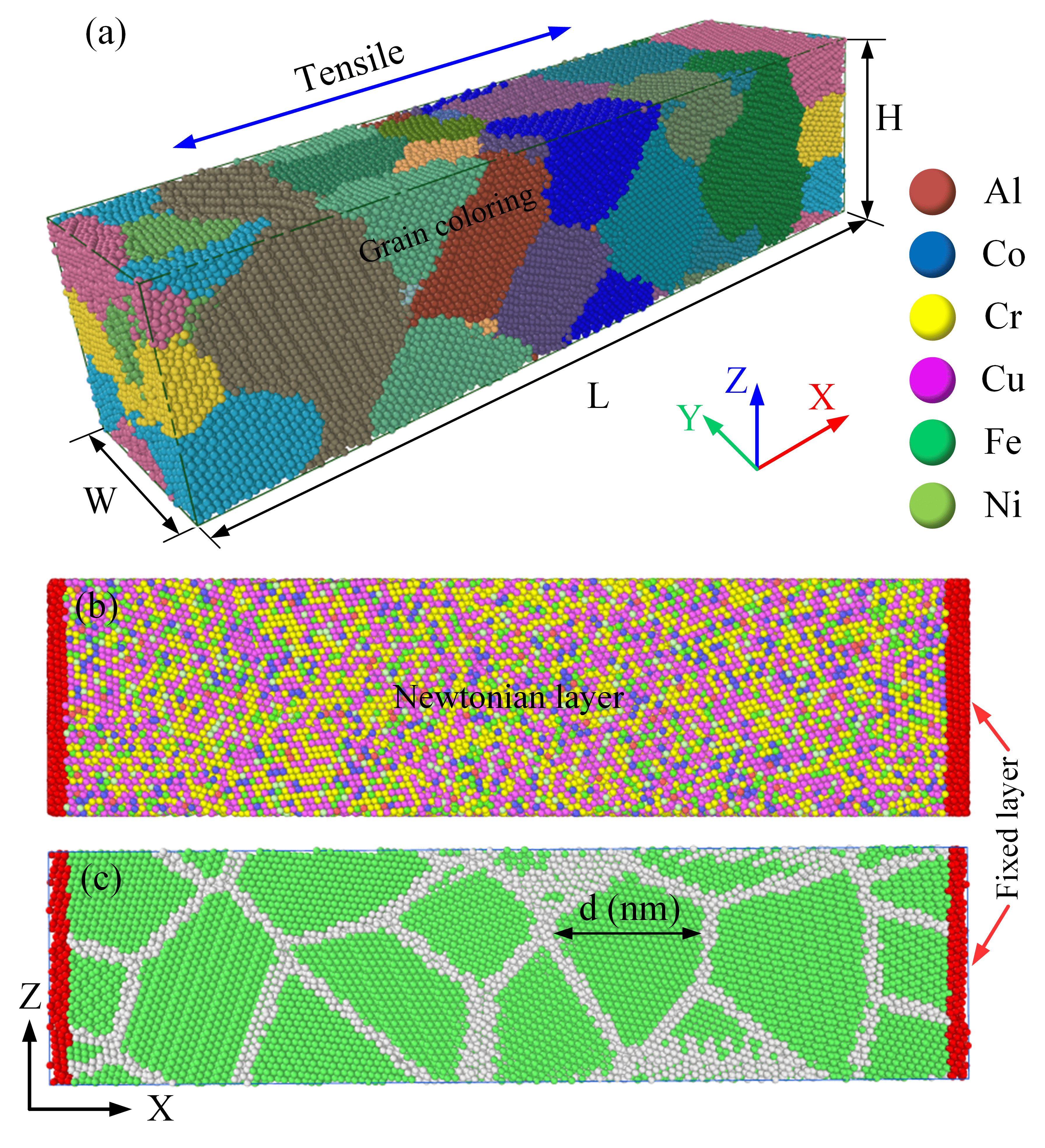}
	\caption{Model of tensile AlCoCrCuFeNi.}
	\label{fig:AlCoCrCuFeNi}
\end{figure}

We subjected the workpiece to uniaxial tensile loading, as depicted in Fig. \ref{fig:AlCoCrCuFeNi}, to reveal the plastic deformation mechanism and mechanical properties of AlCoCrCuFeNi HEA. The results from this dataset are then analyzed. The dimensions of the samples are 300 × 75 × 75 Å$^3$ (L × W × H), and their total atomic count ranges from 146,117 to 146,315 atoms. In the simulated three-dimensional FCC crystal structure, atoms are positioned randomly with displacement coordinates in x, y, and z. The workpiece is aligned in three distinct directions: [1 0 0] along the X-axis, [0 1 0] along the Y-axis, and [0 0 1] along the Z-axis. The elemental chemical compositions, in atomic percent, are approximately 2\% Al, 9\% Co, 32\% Cr, 39\% Cu, 12\% Fe, and 6\% Ni \cite{xie2016molecular, wang2020microstructures, zhu2010microstructures}. 
Periodic boundary conditions were employed in all three directions to establish stable specimen configurations before the tensile loading process. The conjugate gradient algorithm method achieves workpieces in a state of minimal equilibrium energy. Subsequently, samples are equilibrated thermodynamically for 100 picoseconds at ambient temperature and zero pressure using the isothermal-isobaric (NPT) ensemble \cite{nguyen2023cyclic, nguyen2023plastic, sharma2017crystallization}. The velocity-Verlet algorithm is selected with a time step of 2 fs to integrate the motion equation \cite{martys1999velocity, cuendet2007calculation}. In this study, the Nosé-Hoover thermostat and barostat are employed to control temperature and pressure throughout the tensile process \cite{evans1985nose, patra2014nonergodicity}. Table 1 provides details on the workpiece and simulation parameters utilized in this study. Choosing a reliable measure of interatomic potential is essential for obtaining accurate results in MD simulations. Consequently, in order to characterize the interatomic interaction between Al-Co-CrC-u-Fe-Ni, the EAM (Embedded Atom Method) potential was used in this work \cite{daw1993embedded, utt2020grain}.
The effectiveness of EAM potentials has been proven in numerous prior studies involving diverse test processes \cite{xie2013alcocrcufeni}. The total energy Ept is represented as \cite{doan2020influences, zhou2004misfit}:

\begin{align}
    E_{pt} = \sum_{i=1}^{n} E_i = \frac{1}{2} \sum_{\substack{i,j=1 \\ i \neq j}}^{n} \phi_{ij}(r_{ij}) + \sum_{i=1}^{n} F_i(\rho_i)
\end{align}

Where $E_i$ is the atomic potential energy of atom $i$, $\phi_{ij}(r_{ij})$ is the pair energy between atom $i$  and $j$ as a function of their distance $r_{ij}$ and $F_i(\rho_i)$ is the embedding energy term as a function of the local electron density $\rho_i$ at the position of atom $i$. The local electron density can be calculated using: 

\begin{align}
    \rho_i = \sum_{\substack{i,j=1 \\ i \neq j}}^{n} f_{ij}(r_{ij})
\end{align}

Where, $f_{ij}(r_{ij})$ represents the electron density atom $j$ contributes to the particle $i$ site. In the EAM alloy potential model, the pair potentials are defined as:

\begin{align}
    \phi(r) = \frac{A \exp\left[-\alpha\left(\frac{r}{r_{e}}-1\right)\right]}{{1 + \left(\frac{r}{r_{e}}-\kappa\right)^{20}}} - \frac{B \exp\left[-\beta\left(\frac{r}{r_{e}}-1\right)\right]}{
 \left(1 + \left(\frac{r}{r_{e}}-\lambda\right)^{20}\right)}
\end{align}

Where $r_{e}$ is the equilibration spacing between nearest neighbors, $A, B, \alpha, \beta$ are four adjustable parameters, and $\kappa$ and $\lambda$ are two additional parameters for the cutoff. The electron density function is taken with the same form as the attractive term in the pair potential with the same value of $\beta$ and $\lambda$ \cite{zhou2004misfit}. The electron density function is:

\begin{align}
    f(r) = \frac{f_e \exp \left[-\beta \left(\frac{r}{r_e} - 1\right)\right]}{1 + \left(\frac{r}{r_e} - \lambda\right)^{20}}
\end{align}

The pair potential between two different species, for example $a$ and $b$ is then constructed as: 
\begin{align}
    \phi^{nm}(r) = \frac{1}{2} \left[ \frac{f^m(r)}{f^n(r)} \phi^{nn}(r) + \frac{f^n(r)}{f^m(r)} \phi^{mm}(r) \right]
\end{align}

The below equations represent the embedding energy functions, defining the function in different electron density ranges. These equations are \cite{zhou2004misfit}:

\begin{align}
    F(\rho) =  
\sum_{i=0}^{3} F_{ni} \left( \frac{\rho}{0.85\rho_e} - 1 \right)^i, & \rho < 0.85\rho_e 
\end{align}

\begin{align}
    F(\rho) = 
\sum_{i=0}^{3} F_i \left( \frac{\rho}{\rho_e} - 1 \right)^i, & 0.85\rho_e \leq \rho < 1.15\rho_e 
\end{align}

\begin{align}
    F(\rho) =
\sum_{i=0}^{3} F_n \left[ 1-\eta \ln \left( \frac{\rho}{\rho_s} \right) \right] \left( \frac{\rho}{\rho_s} \right)^n, & \rho \geq 1.15\rho_e
\end{align}

Where $F_{ni}$, $F_i$, and $F_n$ are tabulated constants \cite{daw1993embedded, zhou2004misfit, lin2008computational}.

The Morse potential governs the remaining particle interactions \cite{li2016mechanical, li2016atomic}. It is defined as follows:

\begin{align}
    \Phi(u_{ij}) = D \left[ e^{-2\alpha(u_{ij}-u_0)} - 2e^{-\alpha(u_{ij}-u_0)} \right]
\end{align}

Where $\alpha$ and $D$ are constants of reciprocal distance and energy, respectively. $u_0$ and $u_{ij}$ are the equilibrium and the instantaneous distances of the approach of the two atoms, respectively.

The average grain size $d$ is estimated as \cite{zhang2018competing, schiotz1998softening, mendelson1969average}:

\begin{align}
    d = \sqrt[3]{\frac{6V}{N}}
\end{align}

Where $V$ is the total volume of polycrystalline AlCoCrCuFeNi HEA, and $N$ is the number of grains. Eight workpieces with average grain sizes of 12.84, 10.19, 8.90, 8.09, 7.51, 7.07, 6.71, and 6.42 nm with grains of samples 5, 10, 15, 20, 25, 30, 35, and 40, respectively, are used for this study.
The deformation behavior and the structural evolution are exposed in this study using the following analytical methods: (1) von Mises shear strain is shown to analyze the progression of deformation during the tensile process \cite{pham2020structural}. (2) The common neighbor analysis (CNA) method is employed to observe the crystal structure, such as stacking faults and phase transformation, which Honeycutt and Andersen \ developed cite{honeycutt1987molecular}. (3) Dislocation extraction analysis (DXA) is used to reveal dislocation growth during the tension process \cite{qi2021atomistic}. All simulations are carried out using the Large-scale Atomic/ Molecular Massively Parallel Simulator (LAMMPS) \cite{li2016mechanical, wang2017investigation}. Visualization and statistical analysis of the structure are performed using OVITO software \cite{stukowski2009visualization}.

In this section, we investigate the effects of strain rates, temperatures, and grain sizes on the mechanical characteristics of AlCoCrCuFeNi HEA during tensile operations using molecular dynamic simulations. Fig. 2 shows the stress-strain relationship for AlCoCrCuFeNi HEA. The findings show that as temperatures drop, the stress in HEA samples rises. On the other hand, stress is directly correlated with both grain size and pulling speed. This is explained by the fact that in low-temperature samples, where most atoms oscillate close to their equilibrium positions, atoms' thermal mobility is comparatively restricted. High-temperature softening properties are observed as temperatures rise because of the atoms' higher kinetic energy, which decreases atomic interaction and atom-atom bond strength [14, 29].
The phase transformation of HEA samples at various temperatures is shown in Fig. 3, where the CNA method is used to color-code the atoms. All HEA specimens develop dislocations and stacking faults (SF) as a result of the elastic energy that has been stored being released during tension tests [14, 20]. The phase transition of HEA specimens under ultimate stress is displayed in Fig. 3a. Fig. 3b shows a higher prevalence of the HCP structure in the polycrystal HEA sample with a rise in strain value (0.2). The findings show that as simulation temperature rises, the number of unknown structures significantly increases. Simultaneously, the amorphous structure noticeably rises as the samples' HCP structure falls. This behavior can be explained by the fact that higher temperatures, as seen in Fig. 3b, increase the kinetic energy of atoms, magnifying thermal motion and increasing the amorphous appearance at grain borders. As a result, at grain boundaries, the fraction of amorphous structure rises with temperature.
The dislocation distribution of HEA samples, as determined by DXA analysis, is shown in Fig. 4 at various temperatures during tension simulation. The degree of dislocation movement determines how flexible metal materials are [28-30]. Meanwhile, the ability to move in dislocations affects the strength of HEA samples. Fig. 4a shows the dislocation distribution in the HEA sample at ultimate stress. Because of the impacts of workpiece production and annealing processes on the occurrence of GB dislocations in the manufactured HEA samples, dislocations are initially concentrated mostly at the GB. As the temperature rises, Figure 4b shows that all specimens with a strain value of 0.2 exhibits a large number of Shockley partial dislocations, though their frequency decreases. This implies that, with more robust dislocation nucleation at lower temperatures, lower temperatures increase the efficacy of slip behavior generated by Shockley partial dislocations. Because dislocation emission and nucleation at lower temperatures demand a greater external force, the average flow stress and ultimate strength at lower temperatures are greater than those at higher temperatures. The results show that high temperatures hinder the emission and propagation of dislocations. Amorphization inhibits the sliding activation system, as Fig. 4b shows, and this impact becomes more noticeable at higher temperatures. This finding emphasizes how the amorphous structural barrier becomes more prominent as temperature rises [57], which lowers dislocation at higher temperatures.
\begin{figure}[!htp]
	\centering
	\includegraphics[scale=0.12]{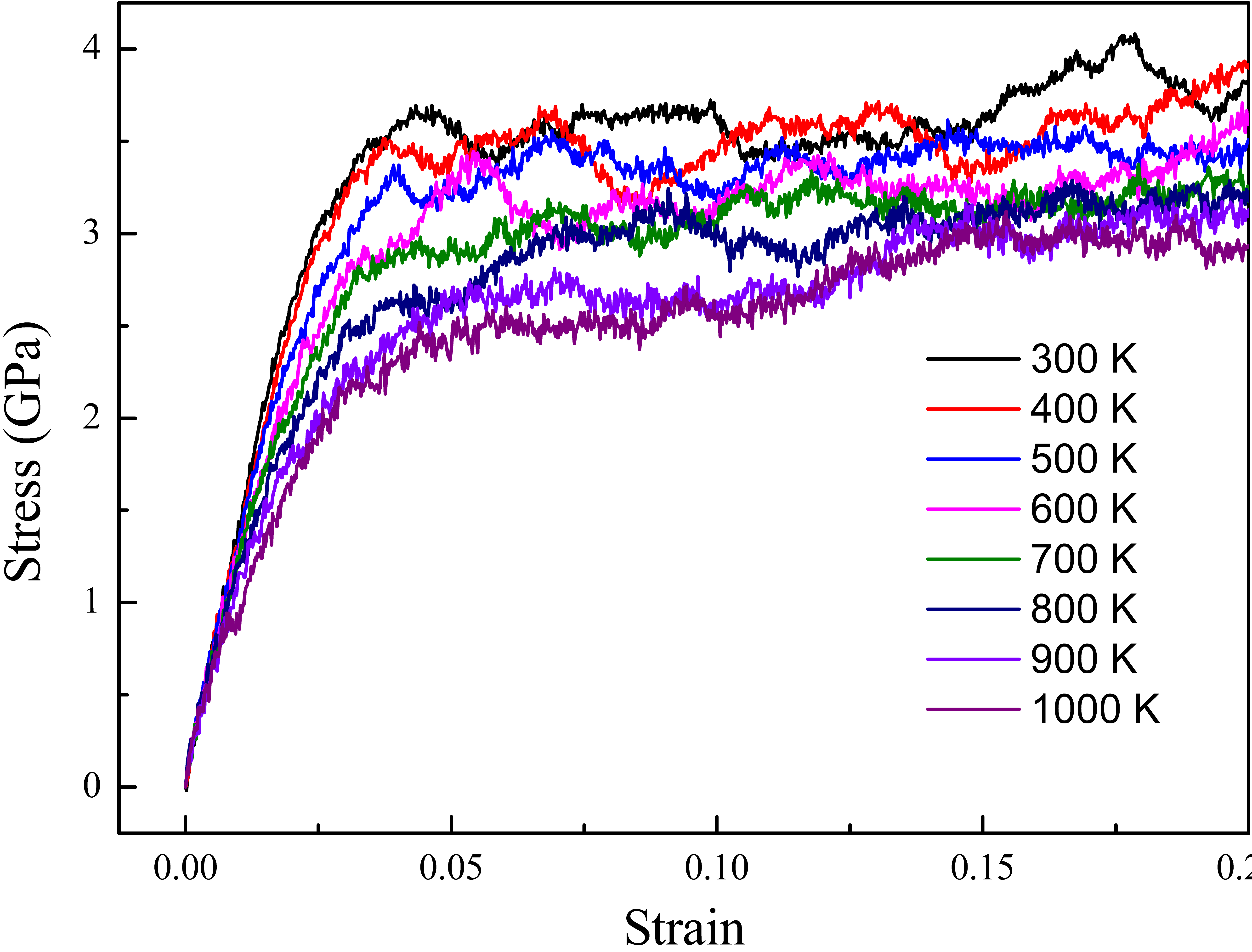}
	\caption{Stress strain.}
	\label{fig:str_strain}
\end{figure}

\begin{figure*}[!htp]
	\centering
	\includegraphics[scale=0.0625]{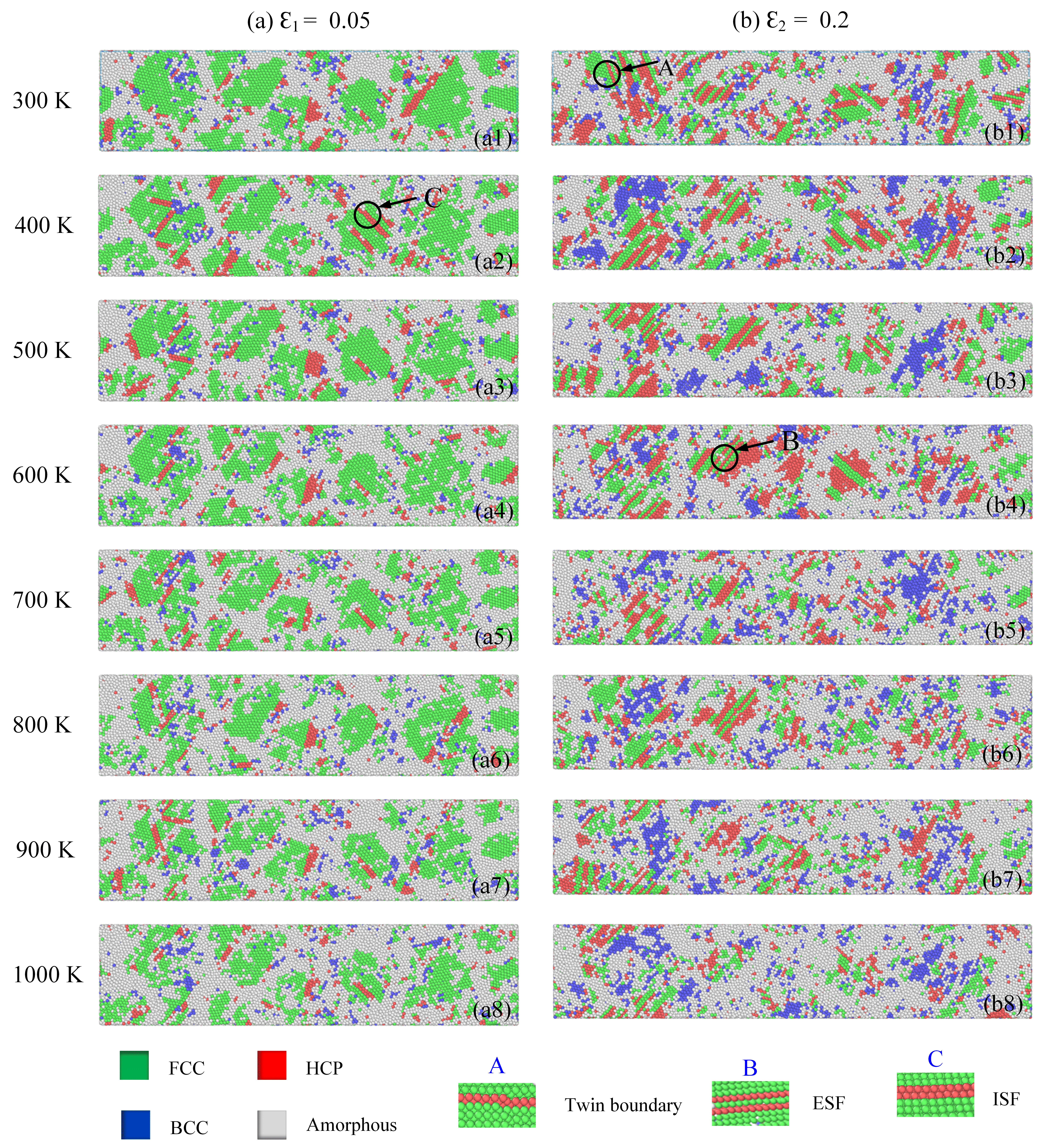}
	\caption{CNA temperature.}
	\label{fig:cna_temp}
\end{figure*}

\begin{figure*}[!htp]
	\centering
	\includegraphics[scale=0.0625]{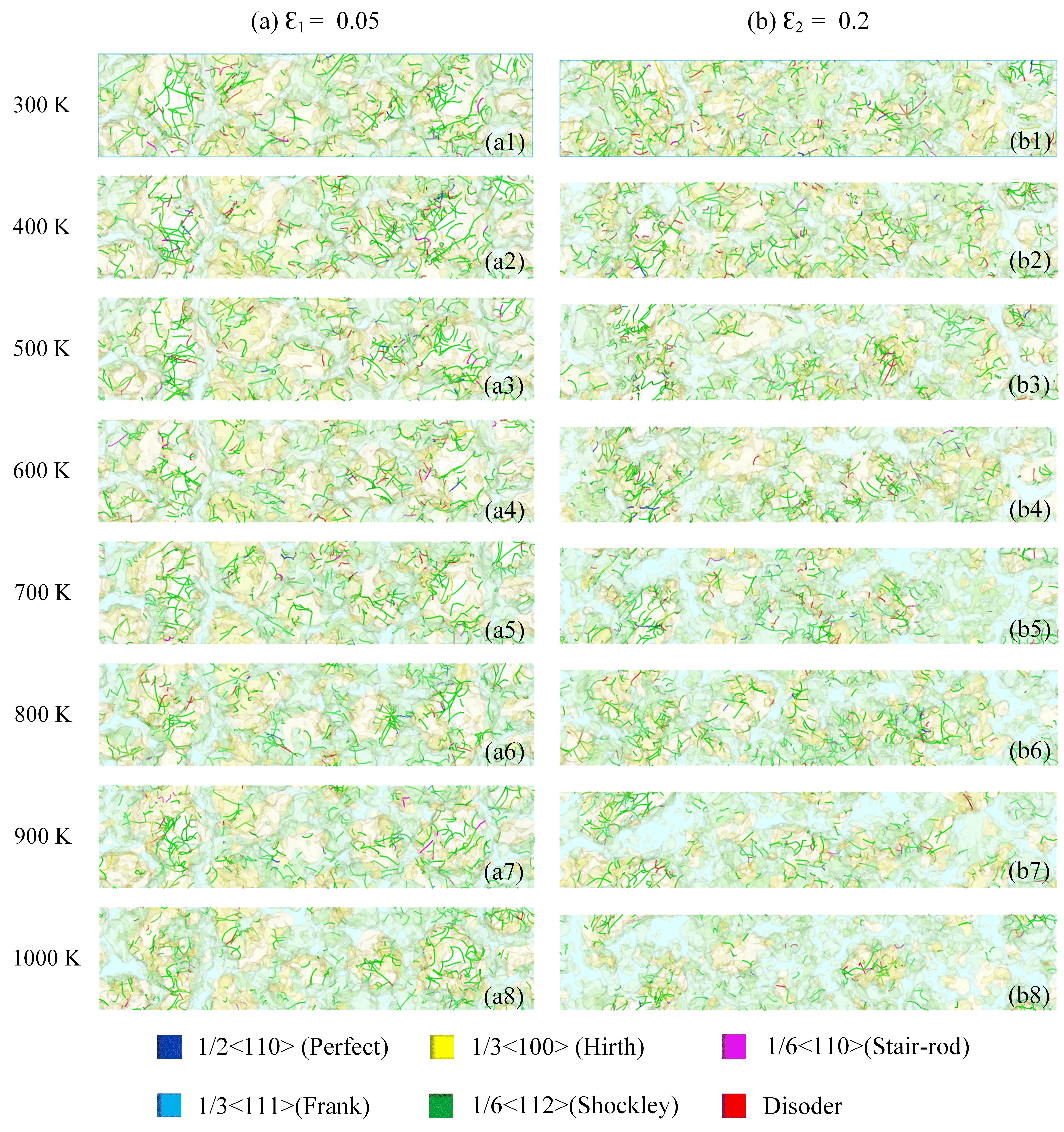}
	\caption{Temperature compress.}
	\label{fig:temp_compress}
\end{figure*}

Finally, we have the statistical analysis Table \ref{tab:data_statis} summarizes the properties of a dataset encompassing 22,021 observations across five parameters: Strain, Temperature, Grain size, Strain rate, and Output. The strain has a fairly uniform distribution with a mean close to the median, ranging from 0 to 0.2. The temperature shows a wide spread, averaging 427.28 K, but with half the values at the minimum of 300 K, suggesting a lower-bound skew. Grain size exhibits less variability, with most data points clustered around the mean of 29.1, and the majority at 20. Strain rate reveals a vast range and high variability, with an average of 2.4e+09, suggesting potential outliers or a heavy-tailed distribution. Lastly, the output has an average of 3.35 and a moderately wide spread, with values from -0.017 to 5.75. The quartile data for both temperature and grain size indicate a significant number of observations at their lower limits, while the strain rate’s large standard deviation points to a diverse range of values. Collectively, these statistics suggest the data exhibits a mix of uniform and skewed distributions, with certain parameters showing potential boundary effects or outliers.

\begin{table*}[]
\caption{Statistical Analysis for Data Values}
\centering
\label{tab:data_statis}
\begin{tabular}{l|l|l|l|l|l|}
\cline{2-6}
{ \textbf{}} &
  { \textbf{Strain }} &
  { \textbf{Temperature (K)}} &
  { \textbf{Grain number}} &
  { \textbf{Strain rate (s-1)}} &
  { \textbf{Stress (GPa)}} \\ \hline
\multicolumn{1}{|l|}{{ count}} &
  { 22021} &
  { 22021} &
  { 22021} &
  { 22021} &
  { 22021} \\ \hline
\multicolumn{1}{|l|}{{ mean}} &
  { 9.9997e-02} &
  { 427.2785} &
  { 29.09913} &
  { 2.400064e+09} &
  { 3.345523} \\ \hline
\multicolumn{1}{|l|}{{ std}} &
  { 5.7791e-02} &
  { 217.8109} &
  { 7.012762} &
  { 4.356896e+09} &
  { 0.901617} \\ \hline
\multicolumn{1}{|l|}{{ min}} &
  { 0 } &
  { 300.0} &
  { 5.0} &
  { 1.000000e+08} &
  { -0.017224} \\ \hline
\multicolumn{1}{|l|}{{ 25\%}} &
  { 5.0000e-02} &
  { 300.0} &
  { 20.0} &
  { 1.000000e+09} &
  { 3.961448} \\ \hline
\multicolumn{1}{|l|}{{ 50\%}} &
  { 1.0000e-01} &
  { 300.0} &
  { 20.0} &
  { 1.000000e+09} &
  { 3.592606} \\ \hline
\multicolumn{1}{|l|}{{ 75\%}} &
  { 1.5000e-01} &
  { 500.0} &
  { 20.0} &
  { 2.000000e+09} &
  { 3.807873} \\ \hline
\multicolumn{1}{|l|}{{ max}} &
  { 2.0000e-01} &
  { 1000.0} &
  { 40.0} &
  { 2.000000e+10} &
  { 5.746938} \\ \hline
\end{tabular}
\end{table*}

\subsection{Machine Learning Algorithms}


Machine learning (ML) enables researchers to analyze large data sets by training models that can be used to classify observations into discrete groups, learn which features determine a metric of performance, or predict the outcome of new experiments. Furthermore, even in fields where such data-intensive methods are not typical, ML can assist researchers in designing experiments to optimize performance or test hypotheses more effectively. From nano-optoelectronics to catalysis to the bionano interface, ML is reshaping how researchers collect, analyze, and interpret their data. These methods will likely evolve into new standards tailored for each field that are complementary to statistics' role in scientific research. In return, nanoscience has the potential to benefit ML by developing electronic or photonic hardware that can implement algorithms more efficiently than conventional computing architectures. Deepening this unique relationship has much to offer both research communities.

\subsubsection{Algorithms function}

\begin{figure}[!htp]
	\centering
	\includegraphics[scale=0.5]{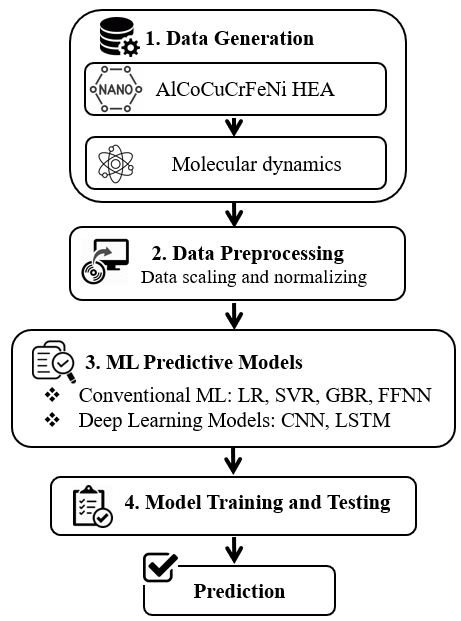}
	\caption{Schematic Workflow of the ML predictive model for stress values}
	\label{fig:workflow}
\end{figure}

This study will focus on applying machine learning techniques for predicting the stress values from the results of data from MD simulation, as shown in Fig. \ref{fig:workflow}.

ML for materials has been developed because of its benefits \cite{jia2021machine}. For example, superconducting material \cite{le2020critical}, which is very expensive to conduct real experiment, and extremely hard to set-up, and conduct the experiments

Espeicially, deep learning is more advanced and robust to materials \cite{masson2023machine}. Therefore, this study we will focus on different machine learning techniques for the regression model from a conventional machine learning to the deep learning models then we can compare the effectiveness of each model based on the evaluation metrics, model complexity, and model computational resource, including:

\begin{enumerate}
    \item Linear Regression (LR): Linear Regression \cite{montgomery2021introduction} assumes a linear relationship between the input features

$\mathbf{x} = (x_1, x_2, x_3, x_4) $, the output $y$. The model predicts $y $ using a linear function:
\begin{align}
    y = \beta_0 + \beta_1 x_1 + \beta_2 x_2 + \beta_3 x_3 + \beta_4 x_4 + \epsilon
\end{align}

where  $\beta_0$ is the $y$-intercept, $\beta_1, \ldots, \beta_4$ are the coefficients for each feature, and$\epsilon$ is the error term.

    \item Support Vector Regression (SVR) \cite{gunn1998support}: SVR aims to find a function $f(x)$ that deviates from $y$  by a value no greater than $\epsilon$ for each training point and at the same time is as flat as possible. Mathematically, it can be represented as:

\begin{align}
    f(\mathbf{x}) = \mathbf{w} \cdot \mathbf{x} + b 
\end{align}

where  $\mathbf{w}$  is the weight vector and  $b$ is the bias. The flatness of  $f$ means minimizing $|\mathbf{w}|$. This is subject to a constraint that for each $i$, either:

\begin{align}
    |y_i - f(x_i)| \leq \epsilon 
\end{align}

or if it is not possible, a slack variable  $\xi_i$ or  $\xi_i^*$ is introduced to soften the margin.

    \item Gradient Boosting Regression (GBR) \cite{friedman2001greedy}: GBR builds an additive model in a forward stage-wise fashion. It allows for the optimization of arbitrary differentiable loss functions. In each stage, a regression tree $h_m(\mathbf{x})$ is fitted on the negative gradient of the given loss function $ L(y, F(\mathbf{x}))$: 
\begin{align}
    F_m(\mathbf{x}) = F_{m-1}(\mathbf{x}) + \gamma_m h_m(\mathbf{x})
\end{align}
 
where $F_m(\mathbf{x})$ is the model up to iteration $m$, and $\gamma_m$ is the step size at iteration $m$.

    \item Feedforward Neural Network (FFNN): FFNN is a feedforward artificial neural network. It consists of at least three layers of nodes: an input layer, a hidden layer, and an output layer. Each node, except for the input nodes, is a neuron that uses a nonlinear activation function. The output $y$  can be represented as:
\begin{align}
    y = f(\mathbf{W}_2 f(\mathbf{W}_1 \mathbf{x} + \mathbf{b}_1) + \mathbf{b}_2)
\end{align}

where $\mathbf{W}_1$ and $\mathbf{W}_2$ are the weights, $\mathbf{b}_1$ and $\mathbf{b}_2$ are the biases, and $f$ is the activation function.

Besides, conventional deep learning has a strong impact on many applications because of its high capacity to learn the complex hidden representation from the data \cite{lecun2015deep}. Therefore, we continue to explore two deep learning models as follows:

    \item Convolution Neural Network (CNN) \cite{albawi2017understanding, li2021survey}: While CNNs are more commonly used for image data, they can also be applied to sequential data. The convolutional layers capture local dependencies, and the fully connected layers predict the output. For a single convolutional layer followed by a fully connected layer, the output can be simplified as: 
    \begin{align}
        y = \mathbf{W}_f \cdot (\text{flatten}(\text{ReLU}(\mathbf{W}_c * \mathbf{x} + \mathbf{b}_c))) + \mathbf{b}_f
    \end{align}
 
where $*$ denotes the convolution operation, $\mathbf{W}_c$ and $\mathbf{b}_c$ are the weights and biases of the convolutional layer, $\mathbf{W}_f$ and $\mathbf{b}_f$ are the weights and biases of the fully connected layer, and ReLU is the activation function.

    \item Long Short-Term Memory (LSTM) \cite{hochreiter1997long, yu2019review}: LSTM  is a type of recurrent neural network (RNN) that can learn order dependence in sequence prediction problems. For a single cell, the output $h_t$ at time $t$ is:
    \begin{align}
        h_t = o_t \odot \tanh(c_t)
    \end{align}
 where $o_t$ is the output gate, $c_t$ is the cell state, and  $\odot$ denotes element-wise multiplication. The cell state is updated through a series of gates that control the flow of information:
 \begin{align}
     f_t &= \sigma(\mathbf{W}_f \cdot [h_{t-1}, x_t] + \mathbf{b}_f) \\
i_t &= \sigma(\mathbf{W}_i \cdot [h_{t-1}, x_t] + \mathbf{b}_i) \\
\tilde{c}_t &= \tanh(\mathbf{W}_c \cdot [h_{t-1}, x_t] + \mathbf{b}_c) \\
c_t &= f_t \odot c_{t-1} + i_t \odot \tilde{c}_t \\
o_t &= \sigma(\mathbf{W}_o \cdot [h_{t-1}, x_t] + \mathbf{b}_o)
 \end{align}

Here, $f_t$ is the forget gate, $i_t$ is the input gate, $\tilde{c}_t$ is the candidate cell state,  $\sigma$ is the sigmoid function, and $\mathbf{W}$ and $\mathbf{b}$ represent weights and biases for each gate, respectively.

\end{enumerate}

\subsubsection{Evaluation Metrics}

Build a regression model, it's crucial to measure its performance using various evaluation metrics. These metrics help understand how well the model has learned and predicts the outcomes. Below are the definitions and mathematical formulas for the metrics 
Mean Absolute Percentage Error (MAPE): 
MAPE is a measure of the prediction accuracy of a forecasting method in statistics. It expresses accuracy as a percentage of the error. It is calculated as:
\begin{align}
    \text{MAPE} = \frac{100\%}{n} \sum_{i=1}^{n} \left| \frac{y_i - y_{\text{pred}, i}}{y_i} \right|
\end{align}

Mean Absolute Error (MAE):
MAE is a measure of errors between paired observations expressing the same phenomenon. It's the average of the absolute errors between the predicted values and the actual values. It is calculated as:
\begin{align}
    \text{MAE} = \frac{1}{n} \sum_{i=1}^{n} \left| y_i - y_{\text{pred}, i} \right|
\end{align}

Root Mean Squared Error (RMSE):
RMSE is a quadratic scoring rule that also measures the average magnitude of the error. It's the square root of the average of squared differences between prediction and actual observation. It is calculated as:
\begin{align}
    \text{RMSE} = \sqrt{\frac{1}{n} \sum_{i=1}^{n} (y_i - y_{\text{pred}, i})^2}
\end{align}

R-squared ($R^2$), also known as the coefficient of determination, is a statistical measure of how close the data are to the fitted regression line. It is also known as the proportion of the variance in the dependent variable that is predictable from the independent variables. It is calculated as:
\begin{align}
    R^2 = 1 - \frac{\sum_{i=1}^{n} (y_i - y_{\text{pred}, i})^2}{\sum_{i=1}^{n} (y_i - \bar{y})^2}
\end{align}

where
\begin{itemize}
  \item \( y_i \) is the true value,
  \item \( y_{\text{pred}, i} \) is the predicted value,
  \item \( \bar{y} \) is the mean of the true values,
  \item \( n \) is the number of observations.
\end{itemize}

\section{Results and Discussion}
\label{sec:result_discussions}

Our experimental setup was executed on the powerful GPUs provided by OpenSource Google Colab’s cloud service. We utilized the scikit-learn library and the Keras framework within a Python environment to implement our models. The data for each experiment was divided into two sets: 70\% for training purposes and 30\% for evaluation to assess the model’s performance.

Drawing insights from prior research on optimizing neural network structures, we adopted strategies to improve our model’s performance and ensure its stability. This included the integration of dropout techniques, with a set probability of 0.25, to prevent overfitting by randomly omitting units during training. Additionally, we chose the GlorotNormal kernel initializer for its efficacy in maintaining a unit’s output variance proportional to its input variance, thereby optimizing the initialization of the network's weights. We also implemented batch normalization to accelerate training and enhance performance by normalizing each layer's inputs. These hyperparameters were carefully chosen and fine-tuned to achieve the best possible model performance.

\begin{table*}[]
\centering
\caption{Detail hyper-parameters summarization for the implementation}
\label{tab:hyper_DL}
\begin{tabular}{l|ccc|}
\cline{2-4}
                                          & \multicolumn{3}{c|}{Models}                                              \\ \hline
\multicolumn{1}{|l|}{Hyperparameters}     & \multicolumn{1}{c|}{FFNN}     & \multicolumn{1}{c|}{CNN}      & LSTM     \\ \hline
\multicolumn{1}{|l|}{Hidden   layers}     & \multicolumn{1}{c|}{3}        & \multicolumn{1}{c|}{4}        & 3        \\ \hline
\multicolumn{1}{|l|}{Number   of neurons} & \multicolumn{1}{c|}{400}      & \multicolumn{1}{c|}{356}      & 400      \\ \hline
\multicolumn{1}{|l|}{Batch   size}        & \multicolumn{1}{c|}{32}       & \multicolumn{1}{c|}{32}       & 32       \\ \hline
\multicolumn{1}{|l|}{Dropout}             & \multicolumn{1}{c|}{0.25}     & \multicolumn{1}{c|}{0.25}     & 0.25     \\ \hline
\multicolumn{1}{|l|}{Learning   rate}     & \multicolumn{1}{c|}{1.00E-04} & \multicolumn{1}{c|}{1.00E-04} & 1.00E-04 \\ \hline
\multicolumn{1}{|l|}{Optimizer}           & \multicolumn{1}{c|}{Adam}     & \multicolumn{1}{c|}{Adam}     & Adam     \\ \hline
\multicolumn{1}{|l|}{Total   parameters} & \multicolumn{1}{l|}{323201   (1.23 MB)} & \multicolumn{1}{l|}{128161   (500.63 KB)} & \multicolumn{1}{l|}{3206801   (12.23 MB)} \\ \hline
\end{tabular}
\end{table*}

The Table \ref{tab:hyper_DL} provides a detailed comparison of hyperparameters used in three different deep learning models: Feed Forward Neural Network (FFNN), Convolutional Neural Network (CNN), and Long Short Term Memory (LSTM). For the FFNN model, there are 3 hidden layers, 400 neurons per layer, a batch size of 32, a dropout rate of 0.25, a learning rate of 6.00E-04, and it uses the Adam optimizer with a total of 323,201 parameters, amounting to 1.23 MB. The CNN model is configured with 4 hidden layers, 356 neurons, a consistent batch size of 32, a dropout rate of 0.25, a lower learning rate of 1.00E-04, also utilizes the Adam optimizer, and has 128,161 parameters, totaling 500.63 KB. Lastly, the LSTM model mirrors the FFNN in terms of hidden layers and neurons, sharing the same batch size, dropout rate, and optimizer, but has a learning rate of 1.00E-04 and significantly more parameters, totaling 3,206,801, which is approximately 12.23 MB. This summary highlights the distinctive configurations and scales of these models, showcasing their varied approaches to handling neural network tasks.

\begin{table}[!ht]
\caption{Performance Comparision for Machine Learning Predictive Models}
\label{tab:comparison_ML}
\begin{tabular}{|l|l|l|l|l|l|l|}
\hline
Evaluation & LR     & SVR    & GBR   & FFNN  & CNN    & LSTM  \\ \hline
MAPE       & 318.17 & 136.13 & 468.9 & 21.32 & 193.06 & \textbf{16.21} \\ \hline
MAE        & 0.422  & 0.202  & 0.91  & 0.092 & 0.178  & \textbf{0.072} \\ \hline
RMSE       & 0.615  & 0.298  & 1.25  & 0.12  & 0.267  & \textbf{0.095} \\ \hline
R2         & 0.10    & 0.85   & 0.98  & 0.98  & 0.90    & \textbf{0.99}  \\ \hline
\end{tabular}
\end{table}

In the presented Table \ref{tab:comparison_ML}, the performance of various machine learning models is evaluated across four metrics, with the LSTM model demonstrating superior results in each category. It boasts the lowest Mean Absolute Percentage Error (MAPE) at 16.21, indicating the smallest average prediction error in percentage terms compared to other models. The LSTM also leads in Mean Absolute Error (MAE) with a value of 0.072, suggesting its predictions are the closest to the true values. Furthermore, with the lowest Root Mean Square Error (RMSE) of 0.095, the LSTM model's predictions are shown to have the smallest variance from the actual data. Lastly, an R-squared (R²) value of 0.99 for the LSTM model means it explains 99\% of the variance in the dataset, which is the highest among the models compared, indicating an exceptionally good fit to the data. These results collectively highlight the LSTM's robust predictive ability, outperforming other models such as Linear Regression, Support Vector Regression, Gradient Boosting Regressor, Feedforward Neural Network, and Convolutional Neural Network across the board in this specific experiment.

\begin{figure*}[!htp]
	\centering
	\includegraphics[scale=0.75]{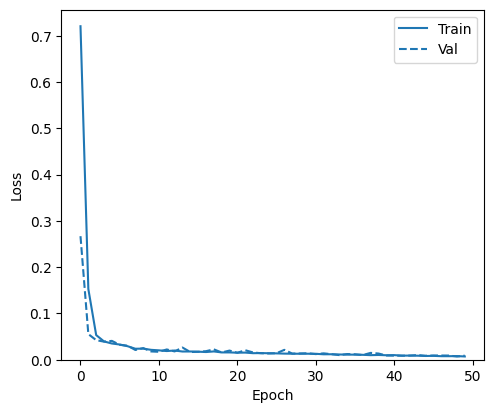}
	\caption{LSTM Learning curve during training (train) and validation (val).}
	\label{fig:lstm_curve}
\end{figure*}

The provided Fig. \ref{fig:lstm_curve} illustrates the loss of an LSTM model throughout 50 epochs of training. Initially, both the training and validation loss decrease sharply, indicating significant learning from the model in the early stages. As the epochs progress, the loss for both datasets levels off, suggesting that the model is converging to an optimal state and that further training yields minimal improvement. Notably, the training and validation loss values remain closely aligned, which is indicative of a well-generalized model that is not overfitting to the training data. The graph also demonstrates a stable and smooth reduction in loss, likely due to a well-chosen learning rate and batch size. The low final loss values for both training and validation suggest that the LSTM has achieved a strong predictive performance on the dataset. However, while the graph indicates a successful training process, additional metrics would be necessary to fully evaluate the model's performance across different aspects such as accuracy and recall, and it would be prudent to further assess the model using a separate test dataset to ensure its efficacy in real-world applications.

\begin{figure*}[!htp]
	\centering
	\includegraphics[scale=0.55]{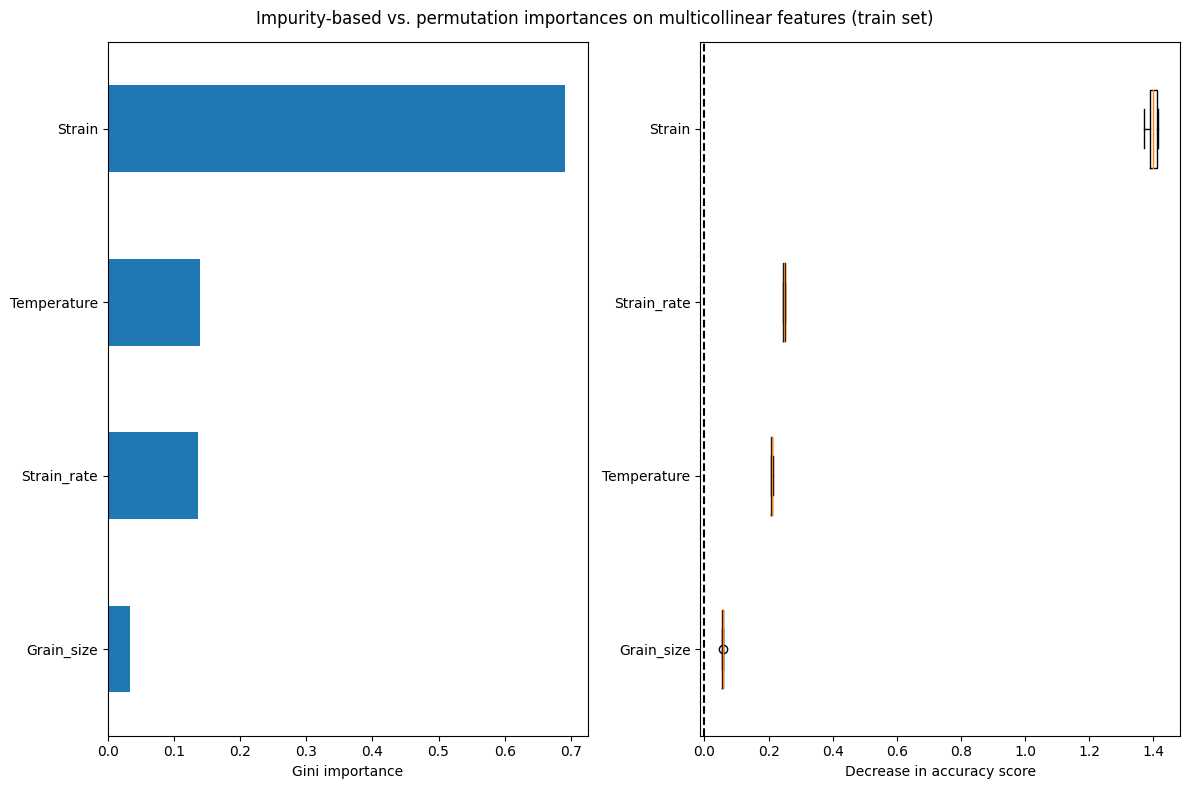}
	\caption{Feature importance regarding to tensile deformation during training}
	\label{fig:feature_importance_train}
\end{figure*}

The Fig. \ref{fig:feature_importance_train} shows two bar charts comparing feature importance from a machine learning model, specifically focusing on multicollinear features within a training dataset. On the left, we have a bar chart titled "Impurity-based vs. permutation importance on multicollinear features (train set)." This chart is displaying the Gini importance of four features: "Strain", "Temperature", "Strain rate", and "Grain size". The Gini importance is a metric used in decision trees and tree ensemble methods like Random Forest to estimate the importance of a feature by measuring how much the tree nodes that use that feature reduce impurity on average (e.g., Gini impurity or entropy). In this chart, "Strain" has the highest Gini importance, followed by "Temperature" and "Strain rate", with "Grain size" having the least importance. On the right side of the image, there is another bar chart with error bars representing the permutation importance of the same features. Permutation importance is calculated by observing how each predictor feature's random reordering (permutation) affects the model performance. The chart shows "Decrease in accuracy score" on the x-axis, which indicates the impact on the model's accuracy when the values of each feature are shuffled. The "Strain" feature has the largest decrease in accuracy, which suggests it has the highest importance based on this method. The error bars indicate the variability or confidence interval of this important measure. Overall, the image suggests an experiment where the importance of features in a predictive model is assessed by two different methods: Gini importance and permutation importance. The experiment likely aims to investigate how multicollinearity (a situation where two or more features are highly correlated) affects these important measures. Notably, "Strain" seems to be the most significant predictor in both methods, while "Grain size" is the least important.

\begin{figure*}[!htp]
	\centering
	\includegraphics[scale=0.65]{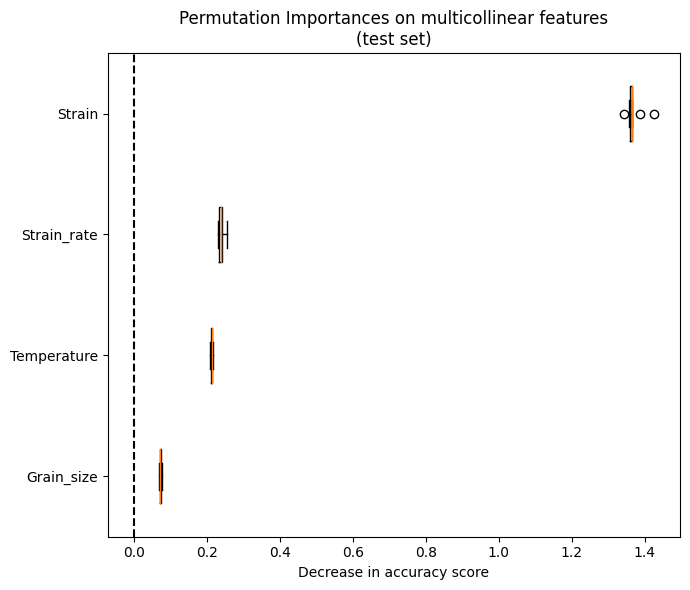}
	\caption{Feature importance regarding tensile deformation during validation}
	\label{fig:feature_importance_test}
\end{figure*}

The Fig. \ref{fig:feature_importance_test} shows the permutation importance of features on a test dataset, presumably following the same methodology as the training set described previously. Permutation importance is calculated by randomly shuffling a feature in the test dataset and determining the change in the model’s accuracy. In this chart, we see the following features listed from top to bottom: "Strain", "Strain rate", "Temperature", and "Grain size". Similar to the previous image for the training set, the x-axis represents the "Decrease in accuracy score", indicating the impact on the model's accuracy when the values of each feature are shuffled. The feature "Strain" has multiple points plotted with error bars, suggesting that the permutation importance was calculated several times, perhaps through a cross-validation process or different test sets, to estimate variability or confidence intervals. The points for "Strain" are scattered with a wide range of values but generally indicate a high decrease in accuracy when this feature is permuted, meaning it is likely an important feature. "Strain rate" and "Temperature" have less variability and show a moderate decrease in accuracy, which points to their lesser but still significant importance. "Grain size" shows the smallest decrease in accuracy, suggesting it has the least importance according to this method on the test set. The spread of the points for "Strain" could indicate that the test sets have varying characteristics or that the feature's importance is highly sensitive to the data it is tested on, a common situation when dealing with multicollinear features. The error bars for each feature give an indication of the reliability of these importance measures; wider bars suggest greater variability in the importance estimate across different test scenarios. Overall, the chart is used to assess the stability of the feature importance across different subsets of data, providing insights into how model performance might generalize to unseen data.

\begin{figure*}[!htp]
	\centering
	\includegraphics[scale=0.5]{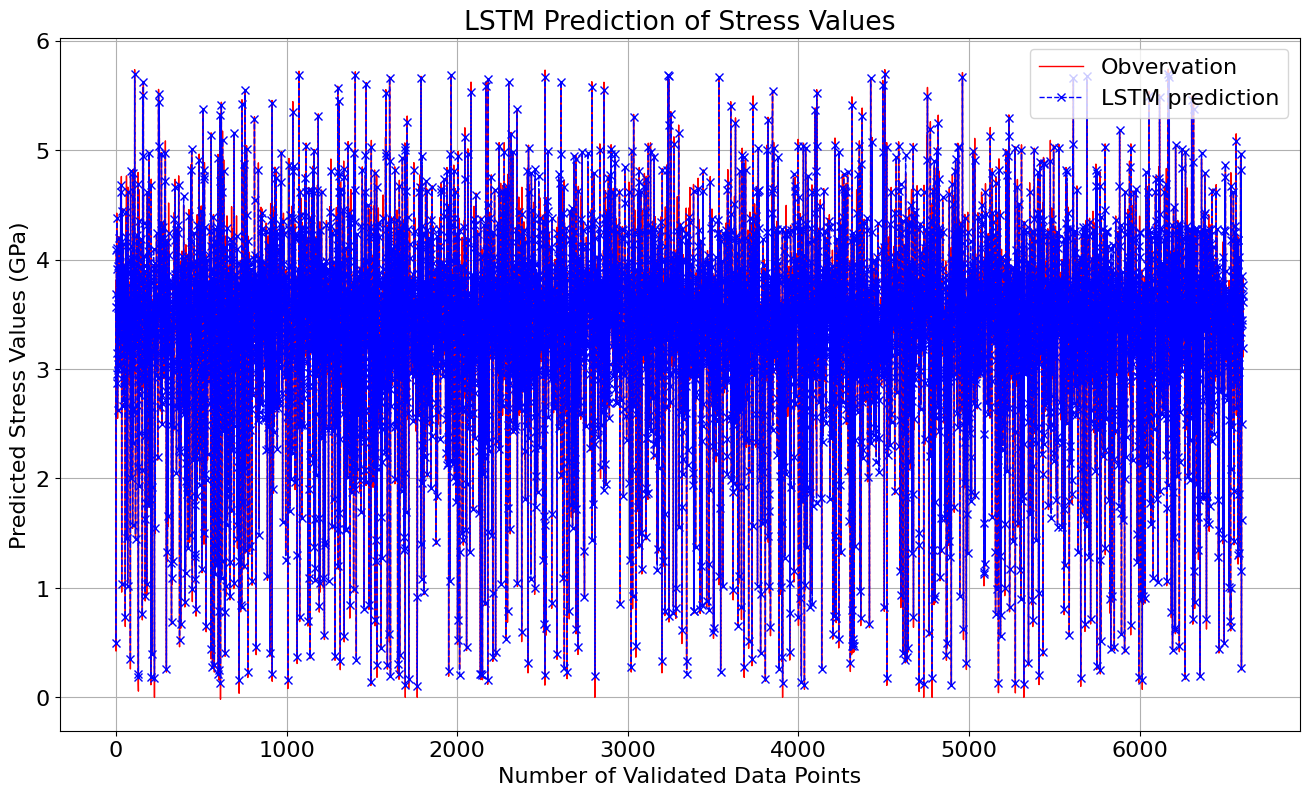}
	\caption{The strain prediction values from LSTM vs real observation}
	\label{fig:lstm_strain_pred}
\end{figure*}

\begin{figure*}[!htp]
	\centering
	\includegraphics[scale=0.5]{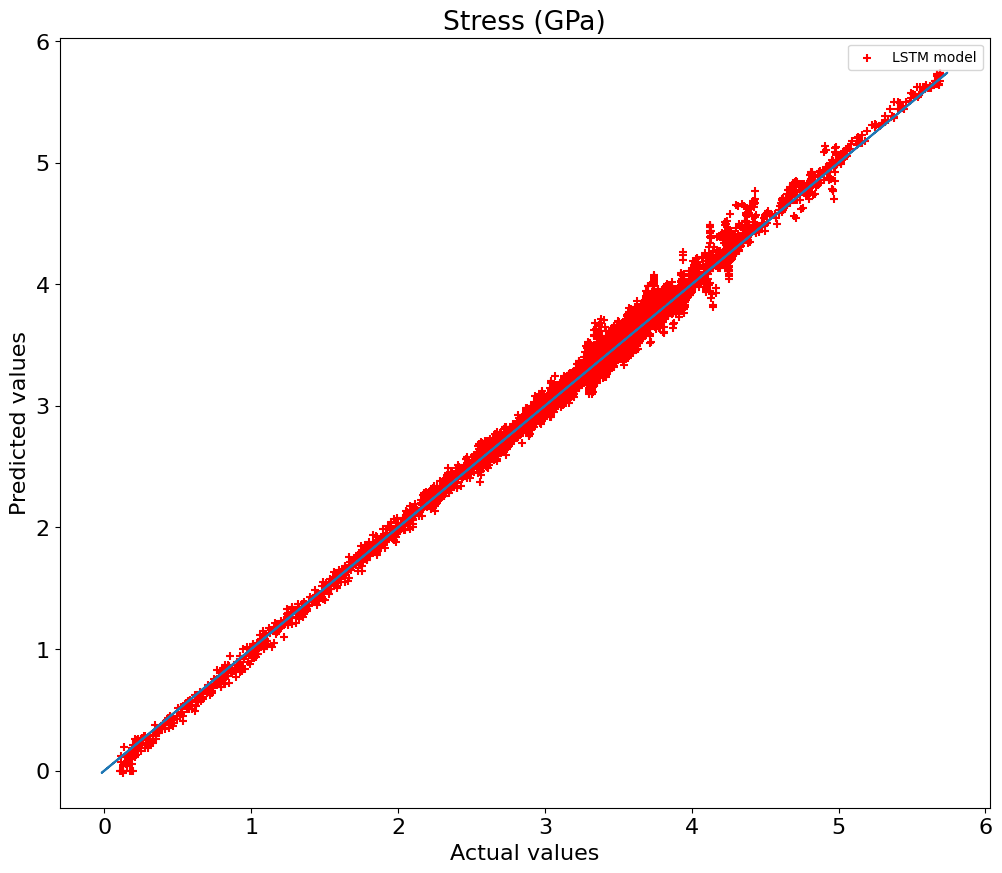}
	\caption{The prediction precision of LSTM compared to actual observation}
	\label{fig:lstm_pred_curve}
\end{figure*}

The Fig. \ref{fig:lstm_pred_curve} you've provided appears to be a scatter plot comparing predicted values from an LSTM (Long Short-Term Memory) model against actual values. The X-axis is labeled "Actual values," and the Y-axis is labeled "Predicted values," suggesting that each point on the graph represents a pair of actual and predicted values for a certain data point. The data points are plotted as red crosses, and there seems to be a blue line that may indicate the ideal situation where the predicted values perfectly match the actual values (a 45-degree line where predicted equals actual). The concentration of red crosses around this blue line indicates that the LSTM model predictions closely align with the actual values. The closer these points are to the blue line, the more accurate the predictions are. The title "Nano strain" could imply that the model is predicting some form of strain at the nanoscale, which could be relevant in fields such as materials science or structural engineering. Overall, the scatter plot suggests that the LSTM model has a good performance, with the majority of predictions falling near the ideal line. However, a more detailed analysis cannot be provided without more context, such as the range of values, units, or specific applications.

Studies on these alloy systems have revealed diverse mechanical properties during tensile testing, influenced by temperature, grain size, and tensile speed factors. The study delves into a comprehensive understanding of high-entropy alloys (HEA). Notably, this research distinguishes itself by integrating tensile deformation into models, utilizing a combination of molecular dynamics simulation and machine learning, setting it apart from previous studies. The tensile strength aligns closely with Afkham et al. [20] findings, and while the temperature and strain rate exceed Elgack et al. [21] results, they are in line with Wang et al. [29] experimental observations. A comparison of mechanical properties across various reports is presented in Tab. \ref{tab:all_compa}, which supports the study's conclusions, reinforcing consistency with previous research. Hopefully, these findings will prove informative and valuable for future investigations into AlCoCrCuFeNi HEA.

\begin{table*}[ht]
\centering
\caption{The comparison of several results between this study and others research.}
\label{tab:all_compa}
\begin{tabular}{|l|l|l|l|l|l|l|}
\hline
\textbf{Materials} & \textbf{Strain rate (s\textsuperscript{-1})} & \textbf{Temperature (K)} & \textbf{Strain (\%)} & \textbf{Stress (GPa)} & \textbf{Method} & \textbf{Reference} \\ \hline
This study & \(10^8 - 2 \times 10^{10}\) & 300 - 1000 & 20 & 3.12 - 5.75 & MD and ML & Bahramyan et al. \cite{bahramyan2020determination} \\
Al\textsubscript{x}CoCrCuFeNi & \(10^{10}\) & 700 - 2200 & 40 & - & MD & Bahramyan et al. \cite{bahramyan2020determination} \\
Al\textsubscript{0.5}CoCrCuFeNi & \(1 \times 10^{-3}\) & - & 7.6 & 1.284 - 1.344 & Experimental & Hemphill et al. \cite{hemphill2012fatigue} \\
AlCrCoFeCuNi & \(10^9 - 10^{11}\) & 300 - 1300 & 40 & 0.6 - 5.5 & MD & Afkham et al. \cite{doan2022effects} \\
FeNiCrCoCu & - & 200 - 400 & - & 0.1367 - 2.158 & MD and ML & Elgack et al. \cite{wu2006adhesive} \\
Al & \(1.7 \times 10^8\) & 300 & - & 2.1 - 4.2 & MD and ML & Mayer et al. \cite{tsai2009deformation} \\
AlCoCrCuFeNi & \(1 \times 10^{-3}\) & 300 - 1473 & 27 & 1.482 - 1.795 & Experimental & Deng et al. \cite{wang2012effects} \\
AlCoCrCuFeNi & - & 300 - 1500 & - & - & MD & Xie et al. \cite{xie2013alcocrcufeni} \\
AlCoCrCuFeNi & \(10^{-4} - 10^{-1}\) & 1073 - 1911 & - & - & Experimental & Shaysultanov et al. \cite{shaysultanov2013phase} \\
CoCrCuFeNiAl\textsubscript{0.5} & \(10^{-4} - 8 \times 10^{-4}\) & RT & 50 & 0.707 - 24.5 & Experimental & Wang et al. \cite{wang2009tensile} \\ \hline
\end{tabular}
\end{table*}

\section{Conclusion}
\label{sec:conclusion}

MD simulations and machine learning are used to examine how the mechanical characteristics and deformation behavior of AlCoCrCuFeNi HEA samples are affected by temperature, tension strain rate, and grain size. The AlCoCrCuFeNi HEA sample softens at high temperatures, which lowers the interatomic linking force. Young's modulus, average flow stress, and ultimate stress are all decreased as a result. In addition, the expansion of the amorphization zone caused by the increase in temperature reduces the total dislocation length.

Additionally, we have implemented different ML algorithms for the predictive model. Experimentally, the results indicate that the Long Short-Term Memory (LSTM) model outperforms all other models in each of the metrics. Specifically, the LSTM achieved the lowest MAPE (16.21), which signifies that it has the smallest average percentage deviation from the actual values, making it highly accurate in percentage error. Additionally, the LSTM model recorded the lowest MAE (0.072) and RMSE (0.095), indicating that on average, its predictions are the closest to the actual values and it has the smallest spread of errors, respectively. Furthermore, the LSTM model achieved the highest R2 score (0.99), which suggests that it is able to explain 99\% of the variance in the target variable, indicating an exceptional level of predictive power and model fit.

In conclusion, the LSTM model not only shows the best performance in each individual metric but does so with a significant margin over other models like Linear Regression (LR), Support Vector Regression (SVR), Gradient Boosting Regression (GBR), Feedforward Neural Network (FFNN), and Convolutional Neural Network (CNN). The superiority of the LSTM model in this comparison suggests that its ability to capture temporal dynamics in the data is particularly effective for the predictive task at hand, making it the preferred choice among the evaluated models.

\section*{Credit Authorship Contribution Statement}
\textbf{Hoang-Giang Nguyen}: Formal analysis, Investigation, Writing – original draft, Visualization, Conceptualization, Writing – Review \& Editing.
\textbf{Thanh-Dung Le}: Algorithm Development, Software Validation, Writing – Review \& Editing.
\textbf{Te-Hua Fang}: Data curation, Funding acquisition, Methodology, Project administration.

\section*{Acknowledgment}

The authors acknowledge the support of the National Science and Technology Council, Taiwan, under grant numbers NSTC 110-2221-E-992-037-MY3.

\bibliographystyle{IEEEtran}
\bibliography{IEEEabrv,Bibliography}
\begin{IEEEbiography}[{\includegraphics[width=1in, height=1.25in, clip, keepaspectratio]{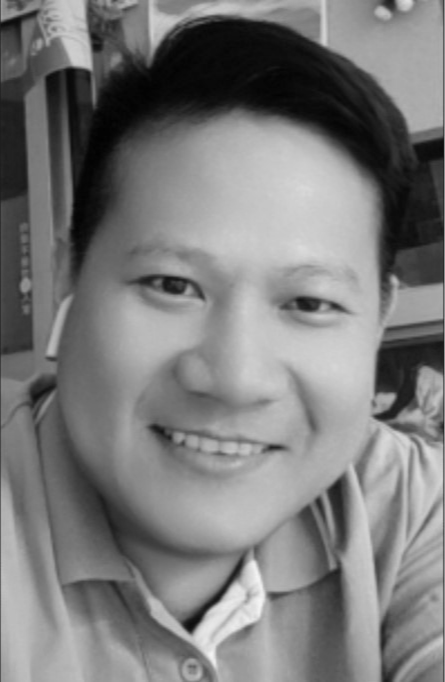}}]{Hoang-Giang Nguyen}  received received a B.S. degree in Mechatronics Engineering and an M.Eng. degree in Automation Control Engineering from Can Tho University, Vietnam. Currently, he is a Ph.D. student in Intelligence Mechanical and Electrical Engineering from National Kaohsiung University of Science and Technology, Taiwan. His main research interests include nanotechnology, molecular dynamics, mechanical properties, machine vision, and intelligent control applications. 
\end{IEEEbiography}

\begin{IEEEbiography}[{\includegraphics[width=1in, height=1.25in, clip, keepaspectratio]{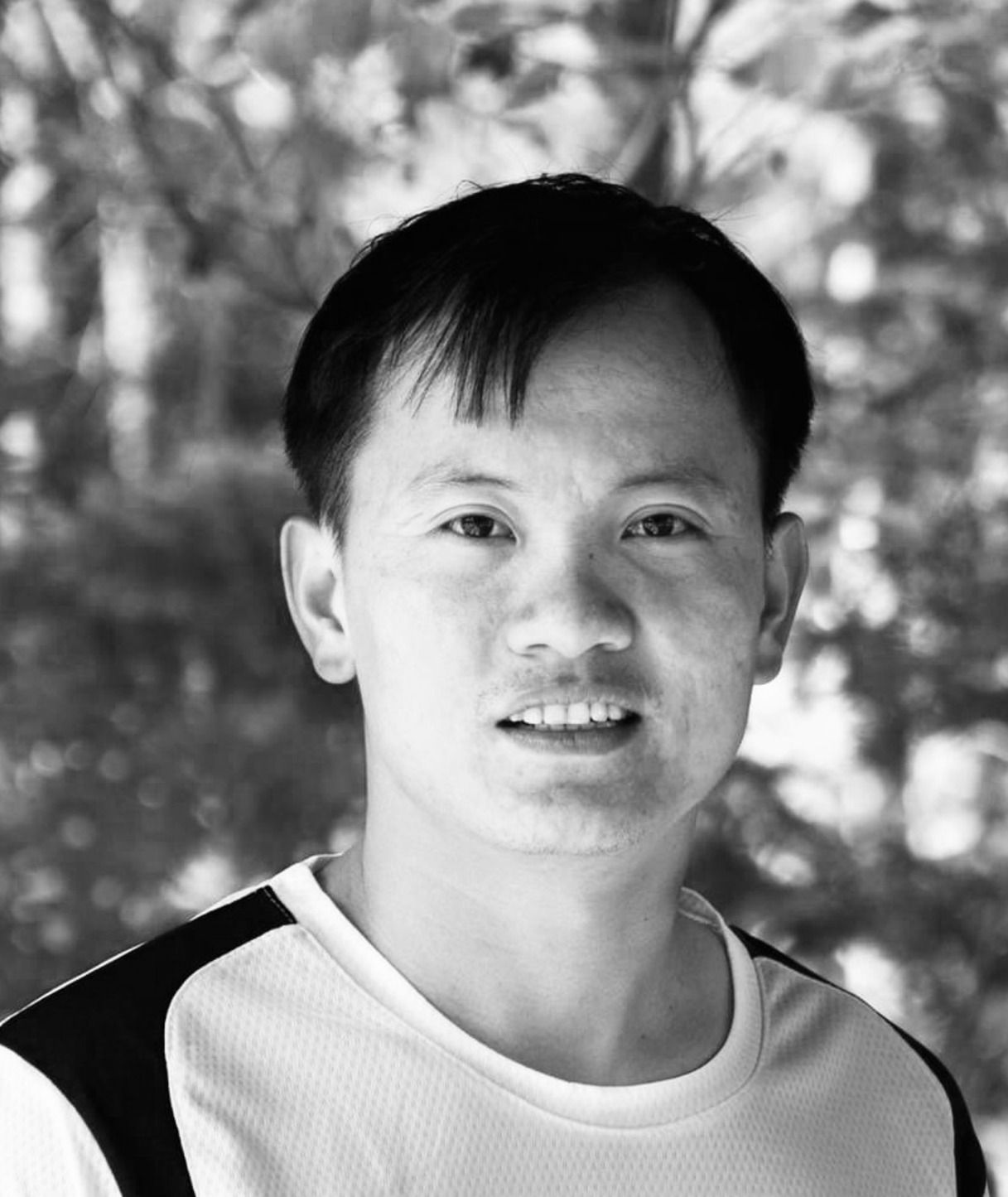}}]{Thanh-Dung Le} (Member, IEEE) received a B.Eng. degree in mechatronics engineering from Can Tho University, Vietnam, an M.Eng. degree in electrical engineering from Jeju National University, S. Korea, and a Ph.D. in electrical engineering from \'{E}cole de Technologie Sup\'{e}rieure (ETS), University of Quebec, Canada. Currently, he is a postdoctoral fellow at the Biomedical Information Processing Laboratory, ETS. His research interests include applied machine learning approaches for critical decision-making systems. Before that, he joined the Institut National de la Recherche Scientifique, Canada, where he researched classification theory and machine learning. He received the merit doctoral scholarship from Le Fonds de Recherche du Quebec Nature et Technologies. He also received the NSERC-PERSWADE fellowship,  Canada, and a graduate scholarship from the Korean National Research Foundation, S. Korea.
\end{IEEEbiography}

\begin{IEEEbiography}[{\includegraphics[width=1in, height=1.25in, clip, keepaspectratio]{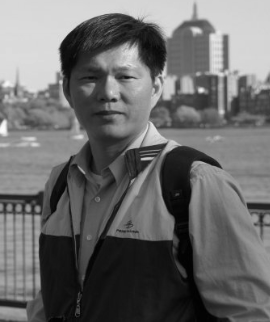}}]{Te-Hua Fang} (IET Fellow, CSME Fellow, ISDS Fellow, TIKI Fellow). He received the M.S. and Ph.D. degrees in mechanical engineering from National Cheng Kung University, Tainan, in 1995 and 2000, respectively. From 2001 to 2005, he was an Assistant Professor and an Associate Professor at the Southern Taiwan University of Technology, Tainan. He joined the Institute of Electromechanical Engineering, National Formosa University, Yunlin, in 2005, as an Associate Professor, where he was a Full Professor in 2007. He joined the National Kaohsiung University of Science and Technology (NKUST), Kaohsiung, Taiwan, as a Full Professor, in 2010. He was Head of the Department of Mechanical Engineering (2018~2022). He is currently the Chair Professor of NKUST and Dean. College of Intelligent Mechanical and Electrical Engineering (2023~now).His current research interests include molecular simulation, mechanical characteristics, and nanotechnology. He is a Fellow of TIKI. He is also a member of ASME. He was a recipient of the Young Outstanding Award and the Outstanding Research Award from the National Science Council, Taiwan, in 2007 and 2011.

\end{IEEEbiography}

\vfill

\end{document}